# Low-Power Optical Actuation of n-GaAs Cantilevers via Surface Piezoelectric Coupling

Ayelén Prado[1], Diego Perez-Morelo[1,2], Santiago Ferreyra[2], Leonardo Salazar Alarcón[1,2] and Hernán Pastoriza[1,2*]

1. *División Dispositivos y Sensores, Gerencia de Física, Centro Atómico Bariloche, CNEA-CONICET, San Carlos de Bariloche, Argentina.*

2. *Instituto Balseiro, Universidad Nacional de Cuyo, San Carlos de Bariloche, Argentina.*

**Abstract**

The mechanical behavior of any semiconductor microstructure is inevitably coupled to light. In the case of micro mechanical resonators (MEMs), carrier generation can affect the quality factor of the resonance through electron-phonon scattering and ohmic losses, but it may also inject energy into the structure and induce movement. Thus, semiconductor MEMs may be regarded as intrinsically optomechanical systems. Here, we report on the optical actuation of a simple-clamped n-GaAs cantilever. This is achieved through modulated nanowatt LED illumination at the resonance frequency. We propose that the mechanism responsible for the coupling between light and movement is piezoelectrically induced stress in the surface depletion layer. Motion may be detected using two methods: by measuring the piezoelectric voltage generated due to deformation or the current arising due to the capacitance changes. In the latter case, a bias voltage must be applied to the device, which leads to nonlinear dynamics. Our results indicate that photothermal and electrostatic effects can be ruled out because i) we measure a very small drift of the resonance frequency with the light modulation offset which, together with finite-element simulations, indicates photothermal effects are not responsible for actuation, ii) we observe a phase shift under different bias voltage polarities, which rules out electrostatic actuation and iii), static measurements performed using an optical profilometer confirm the piezoelectric nature of the interaction with light. In summary, we find homogeneous low power LED illumination to be an effective, simple and convenient method of actuation for piezoelectric semiconductor based MEMs.

**Introduction**

When light reaches a semiconductor surface, several effects occur simultaneously: free carriers are generated and diffuse through the structure, absorbed power heats the system and momentum is transferred due to radiation pressure. These effects can -and have been- exploited to optically actuate semiconductor microstructures [1-8]. Low power consumption, remote free-space access and a simplified fabrication process make optical actuation an attractive alternative for various applications. In this context, $Al_xGa_{1-x}As$ heterostructures have been used to integrate optically active structures within microresonators, enabling a more controlled interaction with light [1-5,8]. In these works, the incorporation of a heterojunction in the cantilever led to the formation of an intrinsic electric field, causing spatial separation of the photo-induced carriers and actuation via the piezoelectric effect. Optical heating and cooling were achieved through the backaction between excitonic absorption and the strain-induced modulation of the band-gap when the resonator was excited by light with a wavelength near the excitonic absorption peak [1]. On the other hand, photo-thermal actuation usually relies on bilayer metallic-semiconductor structures, where the different thermal expansion coefficients between metal and semiconductor lead to bending. In uncoated cantilevers, this mechanism is much less efficient and a thermal gradient must be created across the thickness of the structure. The dependence of the response with actuation position was studied in uncoated $Si_3N_4$ cantilevers by Vassalli et al. [6] where photothermal actuation was shown to be inefficient, generating amplitudes that compete with those due to radiation pressure. Moreover, since thermal relaxation times are slow, thermal actuation is negligible at high frequencies.

A lesser-known effect can be exploited to actuate microstructures: in doped semiconductors, a space charge region (SCR) spontaneously forms between the surface and bulk of the sample [9]. When the sample is illuminated, carrier generation and diffusion modulate this depletion layer. This effect, known as surface photo-voltage (SPV), is highly dependent on illumination wavelength and intensity. Several mechanisms can lead to the modulation of the intrinsic depletion layer in a semiconductor: in doped samples, band bending at the surface due to the occupation of surface states induces photo-generated carriers to spatially separate, changing the surface voltage. Even in intrinsic semiconductors, where the bands remain flat up to the surface, the photo-Dember effect [10] can lead to the generation of a dipole in the sample, due to the variation in absorbed light intensity throughout the sample thickness and the difference in diffusion coefficients between electrons and holes [11]. Once photo-generated carriers are absorbed and diffuse throughout the sample, recombination takes place in different time scales depending on the available transitions. In particular, the (001) surface of n-doped GaAs has been shown to have fast relaxing and slow relaxing SPV components on the picosecond and nanosecond timescales, respectively [12,13]. A longer-lived response, which the authors refer to as a 'piled up' component, was also detected at 300 K. The fast decaying SPV components could enable efficient mechanical excitation of high-frequency flexural modes of free structures of GaAs through the inverse piezoelectric effect [14]. Essentially, light incident on the surface of the free structure modulates the voltage difference between the surface and bulk. This voltage difference leads to a depth-dependent strain in the [110] and [-110] directions, due to the piezoelectric effect, and mechanical bending ensues. In this work, we show that optical actuation can be achieved through this mechanism without a complex optically active structure immersed in the cantilever. Using a modulated LED at low powers (a few nW) we actuate simple-clamped n-GaAs cantilevers. Motion is detected by two mechanisms: either by measuring the voltage difference between the cantilever and substrate, which corresponds to the piezoelectric voltage generated by the deformation, or by measuring the current generated due to the capacitance change between the substrate and cantilever. In the latter case, a DC bias must be applied to the device, which leads to a nonlinear response [15-17]. Our experimental results and simulations suggest that the underlying actuation mechanism is piezoelectric in nature and not related to heating or radiation pressure. To the best of our knowledge, the only previous report of optical actuation of a simple GaAs structure is found in [18], where the vibrations of a polished macroscopical (111) GaAs wafer were excited with light. Direct excitation with a LED instead of a laser source is a cheap, highly effective method for driving resonant semiconductor structures and also a convenient way to eliminate the electrical crosstalk which typically affects measurements with two-port electrical detection and actuation schemes [19].

## Results and discussion

Figure 1(a) shows the setup used for all measurements. The structure used was grown on a (001) n-GaAs substrate using Molecular Beam Epitaxy (MBE) and consists of a 1 μm n-GaAs layer followed by a 4 μm i-$Al_{0.7}Ga_{0.3}As$ sacrificial layer. Simple clamped cantilevers were fabricated either along the [011] or [0-11] directions. Cantilever dimensions are 200 μm x 50 μm x 1 μm, as shown in Figure 1(b). Fabrication details can be found in the materials and methods section. The nominal resonance frequency for the first up-down mode of our structure is around 14 kHz. During all measurements, the sample was placed in a vacuum chamber at a pressure smaller than $10^{-4}$ Torr. Samples were also protected from environmental sources of light, which have a strong effect on vibrations, as we will discuss later. The first up-down mode of our structure could be easily driven by illuminating the cantilever surface with low-power, modulated red LED light (633 nm) at the resonance frequency. As shown in Figure 1(a), there are two possible detection mechanisms: measuring (i) the piezoelectric voltage generated due to deformation or (ii) the current generated by capacitance change. Figures 1(c) and 1(d) show the expected response in each case. Since excitation power is too low to consider photothermal actuation [6], we believe this is a photovoltage modulated piezoelectric effect, as depicted in Figure 1(e). In the case of n-GaAs, surface states are mostly acceptor-like, leading to a depletion of electrons from the space charge region, and upwards band bending. An intrinsic electric field, which points from the bulk to the surface of the sample, is formed. Incident light generates electrons and holes which diffuse to the bulk and surface, respectively. This flattens the bands and lowers the surface voltage ($\phi_s$) and depletion region width. Consequently, the strain in the depletion region changes, leading to the excitation of the up-down mode of the resonator. The bottom plot of Figure 1(e) shows the calculated light intensity as a function of depth. The extinction coefficient for n-GaAs for 630 nm is around 0.198 [20] and, thus, around 87% of transmitted light is absorbed before traveling 0.5 μm. Since barely any light reaches the bottom surface of the cantilever we can focus on the top surface.

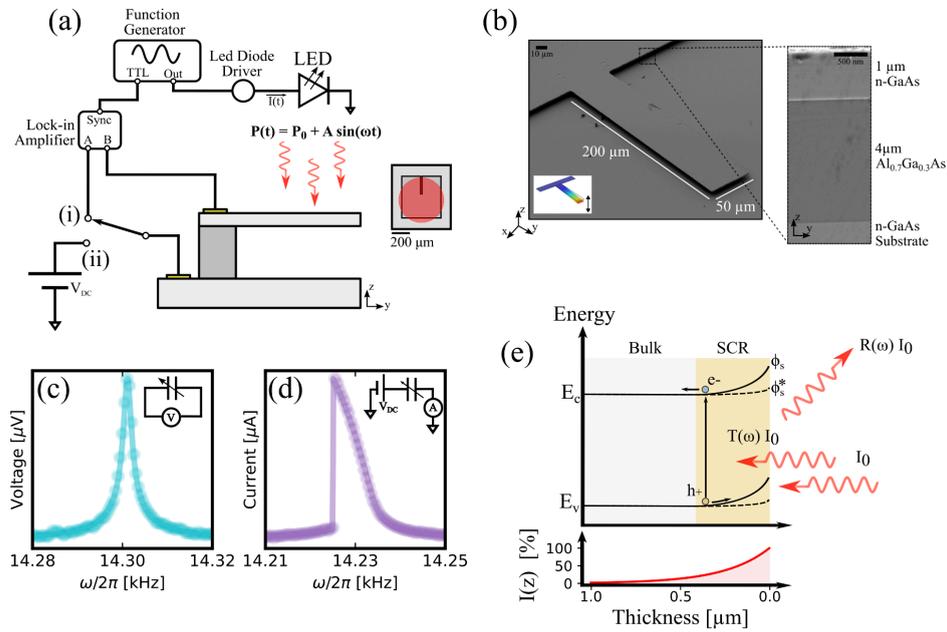

**Figure 1 (a)** Experimental setup for the optical actuation of simple clamped n-GaAs cantilevers. A red (633 nm) LED, driven by a modulated current, is used for actuation. The modulation signal is a sine wave output from a voltage function generator. To avoid the negative part of the signal the function used always has an offset ($P_0$) greater than half the modulation amplitude ($A$). Thus, the light power incident on the cantilever has the form $P(t) = P0 + A sin(\omega t)$. The lock-in-amplifier reference is provided by the digital trigger of the function generator. Two detection configurations were used, as indicated by the switch in the circuit. In (i) motion is detected by measuring the voltage difference between the cantilever and substrate, which is attributed to the piezoelectric voltage generated due to strain. In configuration (ii), a DC voltage ($V_{DC}$) is applied between the cantilever and substrate, and the current induced by capacitance changes during movement is measured. **(b)** A SEM image of the cantilever, the detail of a cross section of the MBE grown layers is shown. The inset shows the finite element simulation of the first up-down mode of the cantilever. Figures **(c)** and **(d)** show the frequency response of the cantilever, when using detection schemes (i) and (ii), respectively. In configuration (ii), a nonlinear response results if the excitation amplitude is high enough. In both cases, the light modulation amplitude was 1.82 W/m² (14.56 nW) with a 0.89 W/m² offset (7.12 nW), and in configuration (ii), a DC bias of 2V was used. Circles correspond to experimental data and solid curves are theoretical fits. Insets show a simplified equivalent circuit for each case **(e)** Actuation: Of the total incident light power, I0, a fraction is transmitted, T(ω)I0, and the remaining reflected, R(ω)I0. Due to acceptor-like surface states a surface charge region (SCR) spontaneously forms in nGaAs. Absorbed photons generate electron-hole pairs that diffuse through the structure, modulating the surface voltage ($\phi_s$) and SCR width. Bottom plot: Light intensity as a function of depth, most of the light is absorbed close to the sample surface.

Figure 2(a) shows the frequency response of the micro resonator for different values of the light modulation amplitude, with the light offset density, $P_0$, set to a constant value of 1.44 W/m² (11.54 nW). Movement was detected using configuration (i), as explained in Figure 1(a). The response is linear, as expected for a simple clamped structure actuated via the piezoelectric effect. No response was found at the second harmonic when the light was modulated at half the resonance frequency. Thus, the excitation can be neither parametric nor electrostatic in nature, since in that case the force between the cantilever and substrate should be proportional to $V^2 \sim \sin(2\omega t)$ [21]. This also rules out electrostriction, which goes as the square of the electric field [22]. In order to show that thermal effects are negligible we studied the behavior of the response with $P_0$. If the cantilever were to heat up homogeneously, we expect its resonance frequency to decrease at a rate of approximately 1 Hz/K (see SI section 1.a). However, when $P_0$ increases from 0.71 W/m$^2$ (5.69 nW) to 5.04 W/m$^2$ (40.38 nW), the resonance frequency decreases less than 0.2 Hz. Considering that (i) heating for a steady light input is larger than what we would expect for pulsed light and (ii) excitation is carried out with modulated amplitudes far smaller than those used in Figure 2(b), we conclude that thermal effects must be negligible. This is also supported by finite element simulations of the thermal dynamics of the system (see SI section 1.b). Figure 2(c) shows other effects of the light offset on the system: On one hand there is a steady decrease in the resonance quality factor ($Q$). On the other, we find an initial increase in amplitude for $P_0 < 2$ W/m$^2$, followed by a decrease as $P_0$ increases. The reduction of the quality factor due to visible light has been reported elsewhere [23], and can be attributed to higher ohmic losses in systems with more free carriers. In contrast, the amplitude variation with light offset has not, to our knowledge, been previously reported. Initially, the light offset seems to inject more energy into the oscillations, but, since the $Q$ factor becomes continuously smaller, the system is clearly less efficient at utilizing it. We have observed the same tendencies when driving the system in a purely electrostatic manner and illuminating with constant light powers. We believe the observed increase in amplitude for low offset powers may be due to an initial filling of unoccupied surface states. This would explain why this effect is seen both when driving the system optically or electrically since the initial curvature of the cantilever changes. This effect will be further studied in future works.

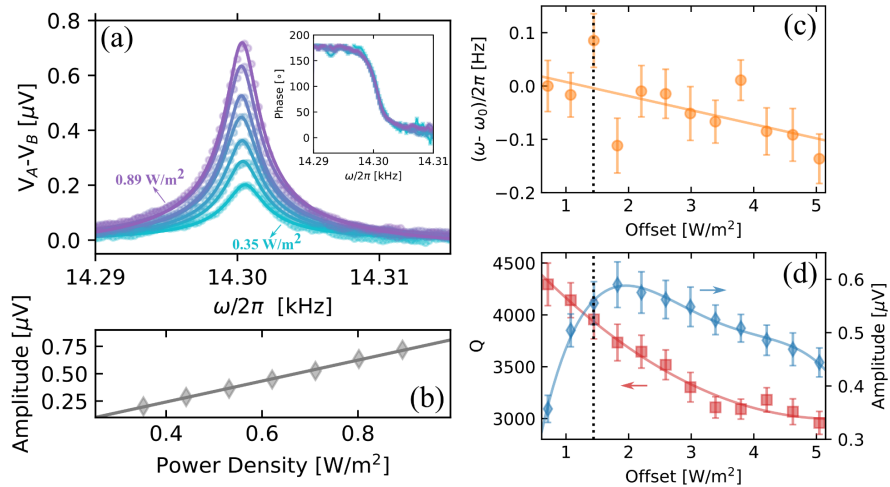

**Figure 2 (a)** Response of the first up-down mode of a simple clamped n-GaAs cantilever. Measurement configuration (i) is used, in which voltage is measured between the cantilever and substrate and no bias voltage is applied, see Figure 1. The light modulation amplitude, $A$, was increased from 0.35 W/m$^2$ to 0.89 W/m$^2$, which corresponds to total incident powers on the cantilever surface of 2.8 nW and 7.2 nW, respectively. $P_0$ remained constant at 1.44 W/m² (11 nW). The inset shows the phase response as a function of frequency, which exhibits the typical behaviour corresponding to a linear resonator. **(b)** Resonance amplitude as a function of $A$. Error bars are of point size. Figures **(c)** and **(d)** show the effects of varying the applied light offset, $P_0$ on the response. In this case light modulation amplitude was 0.89 W/m². As shown in Figure **(c)**, the resonance frequency decreases less than 0.2 Hz when $P_0$ increases from 0.71 W/m² (5.69 nW) to 5.04 W/m² (40.38 nW). Here, $\omega_0$ indicates the resonance frequency measured with the lowest offset power used. **(d)** Effect of incident light offset on the resonance quality factor and amplitude. Lines are guides for the eyes. In (c) and (d) the dotted line indicates the value of $P_0$ used in plot (a).

A nonlinear response is obtained if a constant bias voltage is applied between the cantilever and substrate, as in measurement configuration (ii) (see Figure 1(a)). In this case, an additional term is introduced to the equation of motion of our system, leading to nonlinear dynamics [21]. Figure 3(a) shows the frequency response for light modulation amplitudes between 0.35 and 1.82 W/m² under a bias voltage of 2V. Due to the bias voltage, the amplitude and phase curves show the frequency dependence expected of a duffing oscillator. Interestingly, reversing the polarity of the applied voltage results in a π shift in phase. This is evident both in Figure 3(b) and 3(c) which shows the in phase and quadrature components of the response, and the phase frequency response, respectively, when ±2V are applied between the cantilever and substrate. Notably, this phase shift is not observed when a purely electrical measurement is carried out, as shown in Figures 3(d) and 3(e). In said case, the dominant actuation term at the resonance frequency goes as $V_{DC}V_{AC}$, and cancels out the phase shift in the detected current, which goes as $V_{DC}$. This polarity dependance of the phase response is another indicator which rules out induced electrostatic actuation effects. For more details on the purely electrostatic measurement scheme see SI section 2.

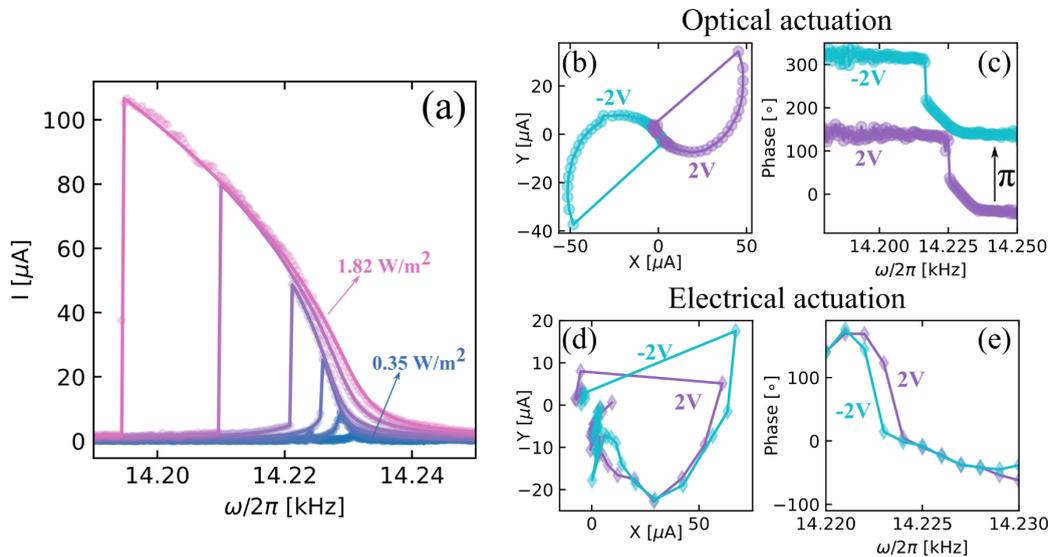

**Figure 3 (a)** Amplitude frequency curves for measurement configuration (ii), see Figure 1(a). The voltage bias, necessary to draw current from the capacitor, also causes nonlinear terms to be added to the equation of motion, resulting in Duffing-like behaviour [17]. Light modulation amplitude values go from 0.35 W/m² (2.82 nW) to 1.82 W/m² (14.58 nW). The bias voltage is 2 V for all measurements. Figures **(b)** and **(c)** show the effect of changing the bias polarity on the phase of the response when actuating the cantilever optically. Figures **(d)** and **(e)** correspond to purely electrostatic measurements and are included for comparison, see SI section 4. In this case, the resonance is driven by modulating the voltage applied between the cantilever and substrate.

The piezoelectric nature of the actuation mechanism can be revealed by static measurements of the cantilever deflection under illumination. The measurements were carried out in air, using an optical profilometer with the Phase Shifting Interferometry (PSI) measurement scheme which is insensitive to light intensity variations, provided there is initially enough light for a measurement to take place [24]. No bias voltage was applied to the device. Figure 4(a) shows the differential change in the cantilever height profile under varying light power densities. These results were obtained for the cantilever used for all previous measurements, which was fabricated along the [0-11] direction. Due to the SCR an electric field is formed which points from the bulk to the sample surface. For a cantilever oriented along [0-11] the resulting surface strain is negative, and the cantilever is initially bent upwards, see SI section 4. Increasing light intensity results in a progressive downward bending of the cantilever. Thus, the tip lowers with increasing illumination. The opposite happens for cantilevers oriented along the [011] direction. Figure 4(b) shows tip deflection measurements made on a single sample in which several cantilevers were fabricated along each crystallographic orientation. With illumination both the surface voltage and SCR width become smaller, and the cantilever loses its initial curvature. The clear distinction between cantilever orientations apparent in Figure 4(b) confirms that the underlying actuation mechanism cannot be heating or radiation pressure, since these effects are not orientation dependent.

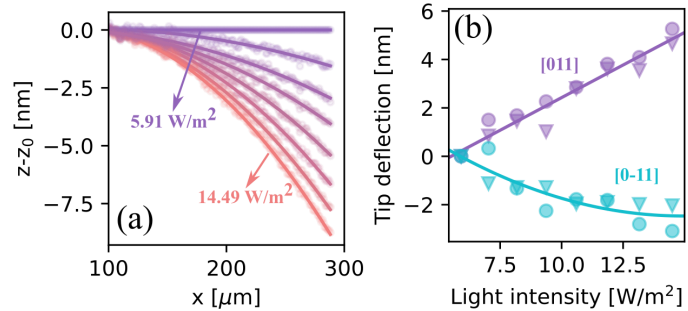

**Figure 4 (a)** Optical profilometer measurements of the cantilever profile for illumination power densities from 5.91 W/m$^2$ to 14.49 W/m$^2$. The profiles shown correspond to the difference with respect to the profile taken with the minimum light necessary to perform a PSI measurement (5.91 W/m$^2$ in this case) **(b)** Tip deflection versus light intensity for cantilevers fabricated in the same wafer. The cantilevers were fabricated either along the [011] or [0-11] directions. Again, the deflection values shown are relative to the one obtained with the minimum light needed for a PSI measurement. Circles and triangles indicate measurements performed on different cantilevers within the same chip.

## Conclusions

The low power LED optical actuation of simple-clamped n-GaAs cantilevers was demonstrated. Efficient actuation was achieved without any optically active structures immersed within the cantilever. When movement was detected by measuring the voltage difference between the top surface and the substrate, linear responses were obtained. The small variation (0.2 Hz) of resonance frequency with light offset shows that photothermal effects are negligible. Application of a constant bias voltage between the cantilever and substrate leads to nonlinear dynamics. Moreover, changing the polarity of the bias voltage shifts the phase response by $\pi$, which is not expected if actuation were due to electrostatic effects. Finally, the piezoelectric nature of actuation with light was confirmed by static measurements carried out in an optical profilometer. The cantilever deflection direction was found to be dependent on orientation, confirming that the interaction with light is piezoelectric in nature. This actuation mechanism is a very effective and cheap way to actuate simple piezoelectric semiconductor structures. In future work we plan to explore the effects of different wavelengths on actuation and check whether high frequency modes can be driven by this method.

**Materials and methods**

Structures were grown on a (001) n-doped GaAs wafer using Molecular Beam Epitaxy (MBE). The device has the following structure: n-GaAs:Si Substrate (n =1-10x$10^{17}$ cm$^{-3}$) / 500 μm, n:GaAs buffer (n= 5x$10^{17}$ cm$^{-3}$) / 0.5 μm, Al$_{0.7}$Ga$_{0.3}$As sacrificial layer/4 μm and n-GaAs device layer (n=5x$10^{17}$ cm$^{-3}$)/1 μm. Simple clamped cantilevers were fabricated either facing the (0 1 1) or (0 1 -1) directions. Cantilever dimensions are 200 μm x 40 μm x 1 μm, and the top structural n-GaAs layer has a nominal doping level of 5 x $10^{17}$ cm$^{-3}$. AuGeNi ohmic contacts [25] were deposited onto the structural layer and the substrate, and flash annealed until linear I-V curves were obtained. After annealing, a bonding layer consisting of a few nm of Cr and a thicker layer of Au was deposited. Cantilever fabrication was achieved by optical lithography to define the structures, followed by chemical etching to eliminate the first GaAs layer ($C_6H_8O_7$:$H_2O_2$ in a 10:1 proportion) and then to selectively remove the sacrificial layer (HF at 1:4 concentration) [26]. Samples were dried in a critical point dryer to avoid collapse due to stiction. The resonator was placed in a vacuum chamber with a pressure of approximately $10^{-4}$ Torr, at room temperature. Purely electrical measurements were carried out by modulating the voltage applied between the cantilever and substrate, as in a typical electrostatic actuation design. Here, actuation was carried out at the resonance frequency (1f) since a bias voltage was applied together with the modulated amplitude. The current generated by capacitance change was used to detect motion. Numerical simulations shown in the Supplementary Information of the thermal dynamics of the cantilever were performed by a commercial finite element-method software. Power density values reported were measured using a power meter and take the spot size into account. For estimates of total power reaching the device, the cantilever surface area was considered.

## Acknowledgements

The authors would like to thank Alejandro Goñi for reading the manuscript and providing useful comments and discussions. Maximiliano Guyón and Axel Bruchhausen for help with measurements and setups. We are also grateful to Matias Lopez for technical help and Reinel Rodriguez for LED spectra measurements. We thank Sebastian Anguiano, Leandro Tosi and Fernando Prado for proofreading the manuscript. Funding was obtained from PICT 201900735 and PICT 202003285.

## Conflicts of interest

The authors declare no conflict of interest.

## Author contributions

AP and DPM contributed to sample processing and device fabrication, conducted the experimental measurements, and analyzed and processed the obtained data, with help from LDS. LDS and AP were responsible for MBE growth of the sample used in the study. HP contributed to the experimental design and provided valuable discussions throughout the project, serving as the project leader. SF participated in preliminary electrical measurements and set-up optimization. The manuscript was written by AP with input from DPM and HP. Finite element simulations shown in the Supplementary Information were carried out by AP. All authors contributed to the revision of the manuscript and approved the final version.